\RequirePackage[2020-02-02]{latexrelease}
\documentclass{aa}
\usepackage{aas_macros}

\usepackage[switch]{lineno}

\RequirePackage[utf8]{inputenc}
\RequirePackage[T1]{fontenc}

\PassOptionsToPackage{dvipsnames}{xcolor} 
\usepackage{graphicx,pstricks}
\usepackage{color}
\graphicspath{{gfx/}}
\usepackage{pgf}

\usepackage{pgfplotstable,filecontents}
\usepackage{booktabs}
\usepackage{url}
\usepackage{xspace}
\usepackage[colorlinks=True,allcolors=RoyalBlue]{hyperref}

\usepackage{natbib}
\bibpunct{(}{)}{;}{a}{,}{,}

\usepackage[nameinlink]{cleveref}

\usepackage{catoptions}

\makeatletter

\def\Autoref#1{%
  \begingroup
  \edef\reserved@a{\cpttrimspaces{#1}}%
  \ifcsndefTF{r@#1}{%
    \xaftercsname{\expandafter\testreftype\@fourthoffive}
      {r@\reserved@a}.\\{#1}%
  }{%
    \ref{#1}%
  }%
  \endgroup
}
\def\testreftype#1.#2\\#3{%
  \ifcsndefTF{#1autorefname}{%
    \def\reserved@a##1##2\@nil{%
      \uppercase{\def\ref@name{##1}}%
      \csn@edef{#1autorefname}{\ref@name##2}%
      \autoref{#3}%
    }%
    \reserved@a#1\@nil
  }{%
    \autoref{#3}%
  }%
}

\usepackage{etoolbox}

\newcommand{\source}[1]{\textsuperscript{\textcolor{blue}{[citation needed]}}\xspace}

\newcommand{\numb}[1]{\textcolor{orange}{#1}}

\renewcommand{\numb}[1]{#1}

\newcommand{\KM}[1]{\textbf{\textcolor{Magenta}{[#1]}}}

\renewcommand{\KM}[1]{#1}
\newcommand{\nbTotBins}{402\xspace}    
\newcommand{\nbTotPairs}{239\xspace}    
\newcommand{\nbBins}{279\xspace}    
\newcommand{\nbPairs}{220\xspace}    
\newcommand{\nbSpecPairs}{36\xspace}    
\newcommand{\nbSpecBins}{127\xspace}    
\newcommand{\nbDiamPairs}{65\xspace}    
\newcommand{\nbDiamBins}{220\xspace}    
\newcommand{\sizea}{0.28\xspace}    
\newcommand{\sizee}{0.2\xspace}    
\newcommand{\sizei}{10.0\xspace}    
\newcommand{\sized}{1.5\xspace}    
\newcommand{\Evlsig}{5.6\xspace}    
\newcommand{\Prmsig}{6.1\xspace}    

\begin{document}

   \title{Deficit of primitive compositions in binary asteroids and pairs\thanks{%
     The catalogs of properties are available
     at the CDS via anonymous ftp to
     \url{http://cdsarc.u-strasbg.fr/} or via
     \url{http://cdsarc.u-strasbg.fr/viz-bin/qcat?J/A+A/xxx/Axxx}}}


   \author{K. Minker
          \inst{1,2}
          \and
          B. Carry\inst{2}
          }

   \institute{Université Côte d'Azur, Observatoire de la Côte d'Azur, CNRS, Laboratoire Lagrange, France\\
              \email{kate.minker@oca.eu}
         \and
             MAUCA — Master track in Astrophysics, Université Côte d’Azur \& Observatoire de la Côte d’Azur, Parc Valrose, 06100 Nice, France
             \\
             }

   \date{Received November xx, 2022; accepted Month day, year}


  \abstract
   {Small binary asteroid systems and pairs are thought to form through fission induced by
   spin up via the Yarkovsky-O'Keefe-Radzievskii-Paddack (YORP) effect.
   This process is expected to depend on their structural strength, hence composition.}
   {We aim to determine how taxonomic classes, used as a proxy for composition, distribute amongst binary asteroids and asteroid pairs compared to the general population.}
   {We compare the distribution of taxonomic classes of binary systems and pairs with
   that of a reference sample of asteroids. We build this sample
   by selecting asteroids to reproduce the orbital and size distribution of the
   binaries and pairs to minimize potential biases between samples.}
   {A strong deficit of primitive compositions (C, B, P, D types)
   among binary asteroids and asteroid pairs
   is identified, as well as a strong excess of asteroids with mafic-silicate rich surface compositions
   (S, Q, V, A types).}
   {Amongst low mass, rapidly rotating asteroids, those with mafic-silicate rich compositions are more
   likely to form multiple asteroid systems than their primitive counterparts.}


   \maketitle
%

\section{Introduction}
\label{sec:intro}

Binary systems have long been of interest to astronomers, as a closely orbiting companion can reveal an objects mass, and subsequently density. This is true for any variety of celestial objects, including stars \citep{1832MNRAS...2...51H,1903PA.....11..240H}, exoplanetary systems \citep{1979Icar...38..136B}, and even black holes \citep{1971Natur.232..465G}.
Of course, asteroids pose no exception to this rule, and the possible discovery of asteroids with companions has been discussed since the early 20th century \citep{asteroidsIIdoasteroidshavesatellitesweidenschilling}. Once the existence of these systems was confirmed with the discovery of Dactyl around (243) Ida in 1993
\citep{1995Natur.374..783C}, the detection and characterization of binary asteroid systems, also known as asteroids with satellites, has been a significant point of interest for observers.


\KM{Several general populations of binary systems are observed. Many of the earliest identified binaries were large ($D_p > 100$\,km) Main-Belt asteroids with small satellites \citep{asteroids3}. A similar population is observed within the Kuiper Belt,  alongside a population of large binary systems with two similarly sized components \citep{2017NatAs...1E..88F}. The largest population of known binary systems is composed of small asteroids in the inner solar system with small satellites, which are the focus of the present study. }

\begin{figure}[t]
\centering
\input{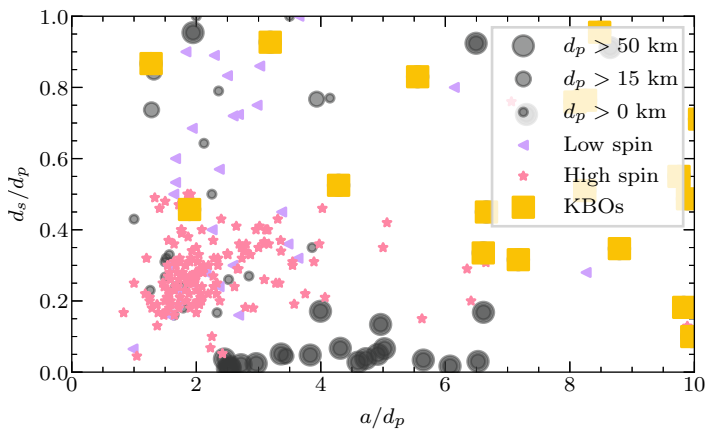}
\caption{%
Known binary systems in the Solar System, including asteroids and Kuiper-belt objects (KBOs). "Low spin" and "High spin" objects, indicated by stars and triangles, are those that have been considered in this study \KM{(objects with $D<15$\,km and $a<2.502$\,au)}. "High spin" objects have a primary rotation period less than \numb{$4$}\,h, and "Low spin" objects have primary rotation periods greater than \numb{$4$}\,h. The diameter of objects that are not included in the study are indicated by marker size, separated into three sets of $d_p >50$\,km, $ 50$\,km $> d_p > 15$\,km, and $d_p < 15$\,km. Diameters $d_s$, $d_p$, and the system semi-major $a$ are in km.
  }
\label{fig:binaryproperties}
\end{figure}

\KM{As of January 10th, 2023}, at least \numb{\nbTotBins} binary systems have been identified,
and an estimated 15\,$\pm$\,4\% of near-Earth asteroids (NEAs)
are expected to be in binary systems \citep{pravec06, asteroids4}, \KM{as are a significant fraction of Main-Belt asteroids.}
A substantial portion of \KM{all} known binary \KM{asteroid} systems share highly
similar properties, containing rapidly rotating primaries
with spin periods close to the stability
limit of \numb{2.4}\,h,
an obliquity approaching $0^\circ$ or $180^\circ$,
a secondary to primary diameter ratio of $d_s/d_p \approx 0.3$,
a primary diameter of less than 15\,km,
and a 2:1 relationship between the system semi-major axis and the primary diameter, as can be seen in \Autoref{fig:binaryproperties}
\citep[or][for a review]{asteroids4}.
\KM{These objects are found throughout the inner solar system, and do not include Kuiper Belt objects (KBOs), and account for over 80\% of known binary asteroid systems.}

These systems are expected to be produced by a common formation mechanism,
a rotational-fission model for which was proposed by \citeauthor{walsh08} in 2008,
with variations proposed by \citet{2011Icar..214..161J} and
\citet{2014ApJ...780...60J}.
Compositional idiosyncrasies of this population of small,
rapidly rotating binary systems are not well studied,
but early models by \cite{walsh08} suggest that dark taxonomic types
would be more likely to form binary systems
as a result of their higher porosity.
An observational study by \citet{pajuelo2018}
found a deficit of C type NEA binaries,
but agreement to the general population within a $1\sigma$
uncertainty for all other types.
More recently, simulations of the structural evolution of
asteroids like Ryugu and Bennu, targets of the
Hayabusa and OSIRIS-REx missions \citep{hayabusafirstarticle,
osirisrexfirstarticlenature},
suggest that dark taxonomic types may in fact be less likely to form binary systems
at high spin rates, undergoing structural deformation rather than rotational
fission under these circumstances \citep{2022NatCo..13.4589Z}.

We aim here to establish whether the distribution of taxonomic classes amongst binary systems
is similar or different than that of the general asteroid population.
In \Autoref{sec:datasets}, we present the data used in this study
and the sources from which it was acquired.
In \Autoref{sec:background}, we discuss how we build a taxonomic reference,
by selecting asteroids from the general population.
In \Autoref{sec:results}, we show the resulting discrepancies in
composition between the binary systems and general population.
\Autoref{sec:discussion} provides a discussion of these results.

\section{Datasets \label{sec:datasets}}

In this study, three main populations were considered.
First, the \numb{\nbTotBins} known binary asteroids.
Second, asteroid pairs, which are expected to have originated
as binary systems that are no longer gravitationally bound
\citep{2008AJ....136..280V}.
There are \numb{\nbTotPairs} pairs of asteroids suspected to share a common origin,
and we mainly consider the primary object of these systems.
Finally, the general population, encompassing all known asteroids for
which reliable data on their dynamics and composition is available.

In order to construct the largest possible dataset,
we included information from a plethora of sources.
Using the \texttt{rocks}\footnote{\url{https://rocks.readthedocs.io/}}
interface for SsODNet\footnote{\url{https://ssp.imcce.fr/webservices/ssodnet/}}
\citep{berthier2022}, we gathered values for physical and dynamical
parameters of all asteroids considered in this study with full traceability
to their original sources, which are outlined in \Autoref{app:data_sources}.

The list of binary asteroid systems we consider was sourced from \cite{johnstonNASA},
complemented with updates from the Website\footnote{\url{https://www.johnstonsarchive.net/astro/asteroidmoons.html}} of the same author, and notices from the Central Bureau for Astronomical Telegrams (CBETs)\footnote{\url{http://www.cbat.eps.harvard.edu/cbet/RecentCBETs.html}}.
The list of pairs was extracted from the recent literature
\citep{2008AJ....136..280V, 2009Icar..204..580P, 2021A&A...655A..14K, 2019Icar..333..429P, 2009AJ....137..111V, 2010Natur.466.1085P, 2017AJ....153..270V, 2016A&A...595A..20Z, 2011MNRAS.412..987R, 2020MNRAS.499.3630H, 2019Icar..333..165M, 2020INASR...5...52K, 2022A&A...664L..17V}.

From the \numb{\nbTotBins} total binaries and \numb{\nbTotPairs} pairs,
a subset of \numb{\nbBins} binaries and \numb{\nbPairs} pairs
were selected that were likely
candidates for formation through the YORP spin mechanism.
Two criteria were used for the selection of these objects:
first, that the diameter of the systems primary $d_p<\numb{15}$\,km,
and second that the system has a semi-major axis $a<\numb{2.502}$\,au for binaries or $a<\numb{2.825}$\,au for pairs,
corresponding to objects within the \numb{3:1} and \numb{5:2} mean-motion resonances
with Jupiter.

The first excludes larger systems which are unlikely to have formed via
the YORP spin mechanism \citep{polishook2011}. The second limits the
set to near-Earth (NEA), Mars-Crossing (MC), Hungaria,
Inner Main Belt (IMB) and Middle Main Belt (MMB) populations.
These \KM{dynamical} populations contain the vast majority of \KM{small} binary systems \KM{($d_p<\numb{15}$\,km)},
as a result of efficient targeted surveys
\citep[e.g.,][]{warnerhungaria1, Warnerhungaria2, Warnerhungaria3}.
Additionnally, these objects are less likely to be influenced by the incompleteness bias
found in the Outer Main Belt (OMB) \citep{marsset2022asteroidbias},
and are closer to the Sun, therefore in a position to be strongly
influenced by the YORP effect \citep{vokro15}. Objects without a known diameter
were not rejected from the set, but the diameter of these objects
was instead estimated as described in \Autoref{sec:background}.

No selection was performed based on the spin period of the primary.
While theoretical models \citep{walsh08,2011Icar..214..161J} suggest
that these systems are formed through rapid rotation, the primary
may have reduced spin today due to losing angular momentum in the
formation of the secondary, or through decelerating YORP
effects \citep{2010Natur.466.1085P}.
Furthermore, easily accessible datasets of spin periods are
typically incomplete, and while early reviews of binary systems
and binary candidates \citep{pravec06} listed typical spin
periods of near-Earth binary systems ranging from
2.2 to 2.8\,h, with outliers trailing to 4h,
current observations show a much broader range \citep{2021-pds-warner}.

Due to limitations in the size of the binary and pair sets,
many taxonomic classes are represented by a low number of systems,
such as the Ch, L and E classes. \KM{This is compounded when separating the samples by dynamical class.}
In order to reduce the effects of statistical uncertainties presented by this,
in some instances, objects are grouped using a reduced taxonomic classification,
where multiple classes are combined \citep[similarly to][]{mahlke2021atlas}.
This is detailed in \Autoref{tab:class}.
Any objects with a sub-class were classified according to their
primary designated; so that an asteroid classified as Sl would be considered an S type.
This is motivated by understandings of known compositional similarities between these
classes, such as those highlighted by \citet{vernazza-beck2016} or
\citet{2022A&A...665A..26M}.

\begin{table}[t]
  \caption{Taxonomic classes grouped as complexes ($\Sigma$),
  with their average albedo from objects in the spectroscopic set, and the number
  of binaries (N$_b$) and pairs (N$_p$) considered here.}
  \centering
\begin{tabular}{llrrr}
\hline\hline
  Classes & $\Sigma$ & $p_V$ & N$_b$ & N$_p$ \\
\hline
  S                 & S & 0.24 & \KM{121} & \KM{76} \\
  Q                 & Q & \KM{0.29} & 10 & 3 \\
  V                 & V & 0.27 & \KM{24} & \KM{13} \\
  C, Ch, B, D, P, Z & C & \KM{0.05} & \KM{14} & \KM{14} \\
  E                 & E & 0.52 & 5 & 1 \\
  K, L, M           & M & 0.17 & \KM{12} & \KM{3} \\
\hline
\end{tabular}

\label{tab:class}
\end{table}

Although Q type asteroids are similar compositions to S type asteroids
\citep{2001MPS...36.1167B, 2011Sci...333.1113N},
they are accounted for separately in the reduced taxonomic
classification due to the previously observed overabundance of
Q type binary and pair systems \citep{Polishook2014}.
\KM{While separating objects by dynamical class could reduce some observational biases, doing so is not practical given the current sample size of binary systems. Instead, a reference sample is selected using the methods described in \Autoref{sec:background}.}


\section{Reference population\label{sec:background}}

The selection of an appropriate reference population is essential
to understanding the surface composition of binary asteroids.
Since a number of observational biases affect the current sample
of known binaries, we aim to mimic this population in order
to minimize the effects of such biases.

\begin{figure*}[ht!]
\centering
\input{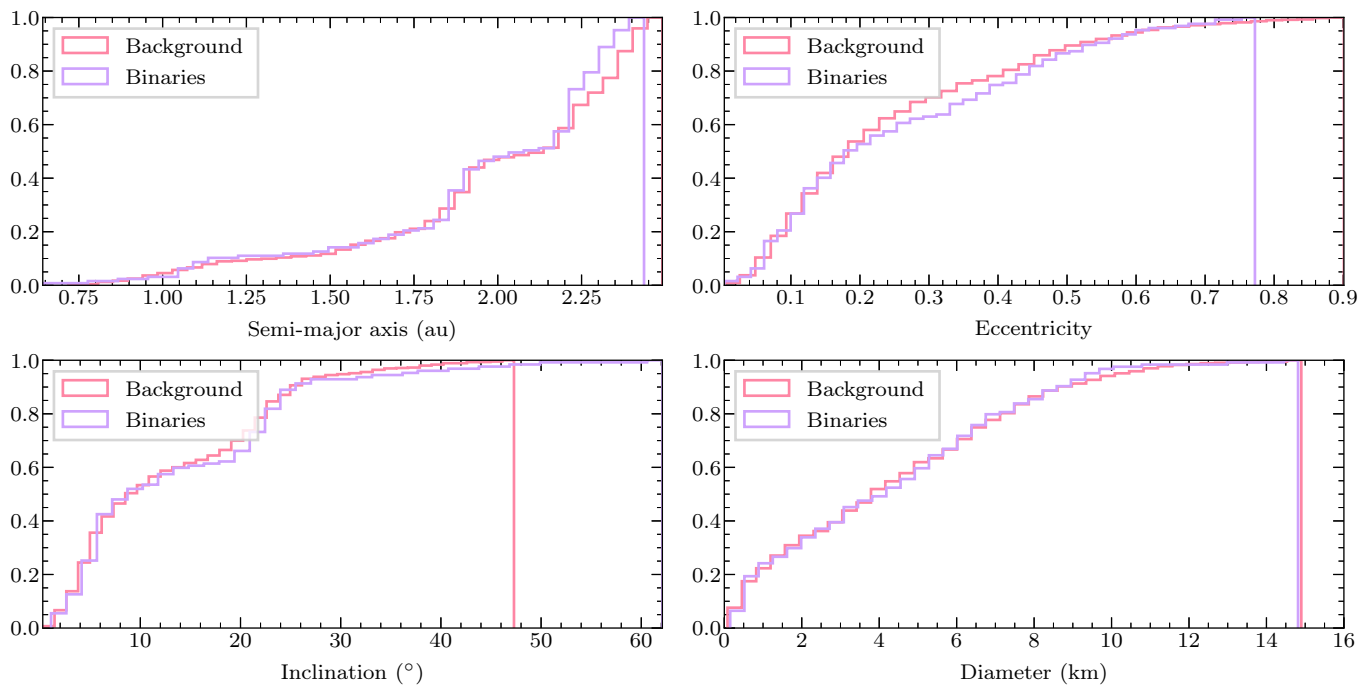}
\caption{%
  Cumulative distributions of the semi-major axis, inclination, eccentricity, and
  diameter of the binary systems (violet) and the selected reference
  population (red).}
\label{fig:binbkgdhistspec}
\end{figure*}

We consider four parameters in the creation of this set; semi-major axis ($a$), eccentricity ($e$), inclination ($i$), and \KM{effective} diameter ($D$).
These parameters are considered for the following reasons. We include the orbital elements ($a, e, i$), as it is well known that there is substantial taxonomic variation as a function of these elements \citep{1982Sci...216.1405G,2010A&A...510A..43C}. The taxonomic distribution of asteroids also shows a substantial dependence on diameter, even when dynamically similar \citep{demeocarry2014}.
Selection across these parameters minimizes potential biases caused
by the over representation of low mass binaries in the population,
as well as observational biases from the ease in observing brighter
taxonomic types \citep{stuartbinzel2004,marsset2022asteroidbias}.

Of the \numb{\nbBins} binaries and \numb{\nbPairs} pairs considered in
the present study,
\numb{\nbDiamBins} binaries and
\numb{\nbDiamPairs} pair primaries have estimates of their \KM{effective} diameters.
We estimate the missing \KM{effective} diameters ($D$) from the asteroid
absolute magnitude $H$ and the average albedo $p_V$ of their taxonomic
class (\Autoref{tab:class}) using \Autoref{eq:diameter}
\citep{1989-AsteroidsII-Bowell}:

\begin{equation}
  D = \frac{1329}{\sqrt{p_V}}  10^{-0.2 H}
  \label{eq:diameter}
\end{equation}

We build the reference set as follows, considering as
an example the set of known binary asteroids for which taxonomy from spectra are available.
We create a four dimensional partition in ($a$, $e$, $i$, $D$) for the set of binaries.
Then, for each object in a box of this partition of space,
$M$ objects were randomly selected with replacement
for the reference set from the total population of asteroids
within the same variable space.
As such, the distribution of selected objects mimic the properties of the binary systems.
$M$ is an integer value selected to maximize the size of the
reference population while maintaining a reasonable number
of instances of object duplication that naturally occurs in boxes
with little taxonomic characterisation.

\Autoref{fig:binbkgdhistspec} presents a comparison of the
properties of the carefully selected reference population with the binary asteroids.
The two distributions are extremely similar.
This suggests that this method is effective in selecting a reference sample
with excellent dynamical similarity to the set of binary systems,
therefore reducing the effects of observational biases present
in the set of binary asteroids.

\begin{figure}[t]
\centering
\input{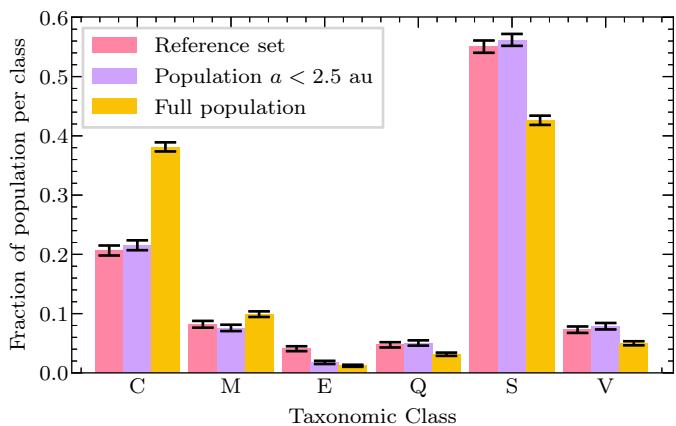}

\caption{
  Illustration of the selection effects on the taxonomic distribution
  of asteroids.
  The reference set \KM{used hereafter} is compared to the
  entire population of asteroids with taxonomy from spectroscopy,
  and those which semi-major axis is less than 2.5\,au.
}

\label{fig:backgroundsbyclass}
\end{figure}

\Autoref{fig:backgroundsbyclass} compares the carefully selected reference sample
using the partitioning method described here with simpler selections.
Most notably, there is a substantially higher abundance of C class
asteroids in the general population of asteroids with spectra than the carefully
selected background.
This discrepancy is slightly alleviated when a restriction on the semi-major
axis of the background population is imposed,
but there is still a strong statistical disagreement.
A similar disagreement can be seen in the S class.
These discrepancies are due to the non random distribution of asteroid
taxonomic classes in the asteroid belt.
This has been known for decades \citep{1982Sci...216.1405G}, with a higher abundance
of primitive types (C/P/D) in the outer belt than in the inner portions of the
belt. More recently, it has also been shown that the relative importance
of taxonomic classes is not constant with diameter
\citep{2013Icar..226..723D,demeocarry2014,2022A&A...665A..83B}.
This demonstrates that it is essential to select an appropriate reference
population in order to minimize the effects of observational biases
affecting both the binary and general populations.

To select an appropriate value of $M$, it is important to understand
the possibility of duplication in the reference sample.
Certain areas of the 4D ($a$, $e$, $i$, $D$) space are significantly
emptier than others. Notably, this includes most of the areas in which
NEAs are found, which means that these zones are at high risk
of excessive duplication when randomly selecting objects in these zones.
This could potentially cause significant biases in the reference set, not only
increasing the likelihood of over-including binaries in the reference,
but could also to extreme over-representation of unusual
taxonomic classes that may be present in the space.
Because of this, the grid used to build the reference sample cannot
be arbitrarily small, and $M$ cannot be arbitrarily large.
A large reference sample is, however, ideal to minimize
statistical uncertainties, so a small value of $M$ is also undesirable.
We examined the duplication rates occurring with several values
of $M$ and different partitionings, and ultimately determined
that a value of
$M=\numb{20}$ and boxes of
\numb{\sizea}\,au,
\numb{\sizee},
\numb{\sizei}\degr, and
\numb{\sized}\,km in ($a$, $e$, $i$, $D$) provided a reasonable balance
between duplication and sample size for this set.

\begin{figure}[t]
\centering
\input{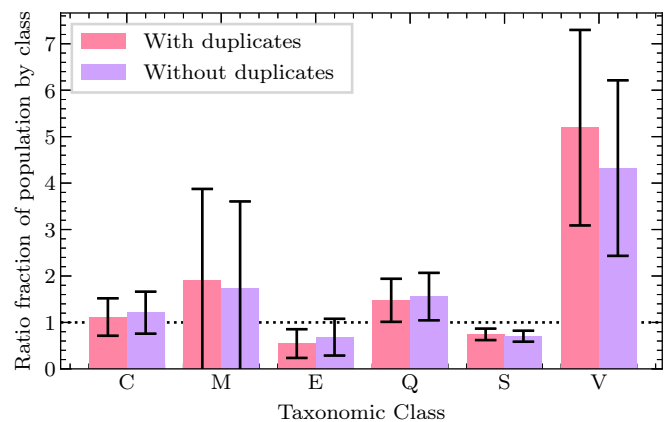}
\caption{Ratio of binary systems to the reference sample, amongst NEAs in the spectroscopic set, allowing for duplicates
  or removing them. The results appear robust against duplication.}
\label{fig:duplicatecomparison}
\end{figure}

The distribution of duplicated objects is not homogeneous within dynamical classes,
largely due to the high prevalence of binary systems within the near-Earth space,
and the low percentage of asteroids belonging to that population.
This inconsistency could cause a bias towards taxonomic types prevalent amongst NEAs,
such as S and Q type asteroids. However, 
the population with
duplicates and without duplicates show a similar taxonomic distribution,
as can be seen in \Autoref{fig:duplicatecomparison}, suggesting that the
bias from the anisotropy of the duplicates is not substantially affecting the distribution of classes.
Amongst NEAs where we see the most duplication, the largest discrepancies between the
set with duplicates included and the set with duplicates excluded occur in the C and V complexes.
Nevertheless, all taxonomic classes show agreement between the sets much below the 1\,$\sigma$ level.

To minimize the effects of the inconsistency between different taxonomic schemes,
the set of binaries was divided into subsets, those with spectroscopic classifications
and those with photometric classifications.
Subset membership is mutually exclusive.
A separate background population was selected for each of these subsets using
the process described above. A similar division was made for the set of asteroid pairs,
considering the taxonomy of the primary of the pair, for a total of four
subsets and four corresponding reference populations
(\Autoref{tab:sample}).
Set membership of asteroids is indicated in \Autoref{app:data_sources}.

\begin{table}[t]
  \caption{$M$ values for the selection of the four reference populations. }
\centering
\begin{tabular}{llrrr}
\hline\hline
  Population & Technique & $M$ \\
\hline
  Binaries & Spectroscopy & 20\\
  Binaries & Photometry   & 30  \\
  Pairs    & Spectroscopy & 15 \\
  Pairs    & Photometry   & 100  \\
\hline
\end{tabular}

\label{tab:sample}
\end{table}


\section{Results\label{sec:results}}

\subsection{Spectroscopic set}

For objects considered in the spectroscopic set, only those
with a taxonomy available derived from the method of \citet{2022A&A...665A..26M}
were considered. The methodology used in this classification is
more technically nuanced than previous studies,
and accounts for both visible and near-infrared spectra.
This classification also accounts for albedo when available,
providing another dimension of information, unlike most recent
spectroscopic taxonomies \citep{bus2002, busdemeotax}.
Most objects with known spectra are included in this classification.
However, only a few thousand objects currently have known
spectra\footnote{Excluding the recent publication of Gaia DR3 visible spectra of 60,000 asteroids \citep{gaia-dr3}},
severely limiting the available reference population of this set.
Since known binary systems are particularly well represented in
the spectroscopic set due to targeted studies by
\cite{pajuelo2018},
this set is comparatively more complete,
with
\numb{\nbSpecBins} out of \numb{\nbBins}
binary systems represented. Pairs are poorly included, with only
\numb{\nbSpecPairs} out of \numb{\nbPairs} pairs represented
\citep{Polishook2014, Duddy2013}.
We compute the relative incidence of binaries and pairs in the general population by
taking the ratio of their taxonomic distribution to that of the reference sample.

We estimate the uncertainties from the statistical error in set membership, as follows:

\begin{equation}
    \sigma=\sqrt{\frac{p (1-p)}{n}}
\label{eq:percentuncert}
\end{equation}

\noindent  where $\sigma$ is the statistical uncertainty, $p$ is the percentage of the set that a given population occupies, and $n$ is the total number of objects in the set.
Standard error propagation techniques then provide the uncertainty in the ratios:

\begin{equation}
  \sigma_r = \left|\frac{x}{y}\right|\sqrt{\left(\frac{\sigma_x}{x}\right)^2+\left(\frac{\sigma_y}{y}\right)^2}
\label{eq:ratiouncert}
\end{equation}

\noindent where $\sigma_r$ is the statistical uncertainty in the ratio, $x$ represents the percentages associated with the population of either binaries or pairs, and $y$ represents the percentages associated with the corresponding background population.

\begin{figure}[t]
\centering

\input{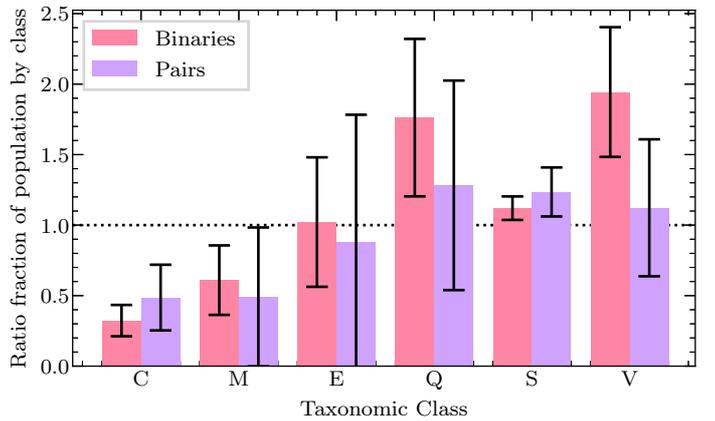}
\caption{Taxonomic distribution of binary asteroids relative
  to the general population, as measured from spectroscopy.
  A ratio below one indicates an
  under representation of the class among binaries, while a ratio
  above one implies that the class is more represented among binaries.}
\label{fig:spectraresults}
\end{figure}

The results of this analysis can be seen in \Autoref{fig:spectraresults}.
There is a significant over-representation of Q and V complex
asteroids amongst binary systems, and a significant under-representation
of C and M complex asteroids.
Amongst pairs, we find a significant under-representation of C and M complex asteroids as well,
and a marginal over representation of S types.

An additional notable finding within this set is the extremely high prevalence
of binary systems amongst V type NEAs, with \numb{$25 \pm 8 \%$}
of known V types with associated spectra being known binary systems
(7 out of a total of 28). While in agreement with the expected total
fraction of binary systems amongst NEAs ($15\pm 4\%$, \cite{pravec06}),
this number is unexpectedly high considering the fact that there have been no obvious
campaigns to search for binary systems amongst this population and that
there is no reason to assume completeness of the true set of binary systems amongst the currently known population,
especially considering the recent discovery of a satellite orbiting V type (7889) 1994 LX in September 2022.

As visible in \Autoref{tab:VtypeNEA}, these binaries were discovered by different researchers,
both from radar and lightcurve measurements.
Since little to no information is available on objects that have been similarly investigated,it is difficult to conclude how many V type NEAs have been searched for satellites without detection.
Some of these objects were specifically targeted for spectroscopic studies due to their  status as binary systems
\citep[such as (348400) 2005 JF21, targeted by][]{pajuelo2018}.
Still, the high fraction of binary systems that are V types, combined with the high fraction
of V type NEAs that are known binaries suggests that there is a striking correlation between the two.

\begin{table}[ht]
  \caption{Discovery information for V type binary NEAs. Asteroid name and number (Asteroid), year of announcement (Year), discovery technique (Technique), either lightcurve (LC) or radar (Radar), and primary investigator of the discovery (PI).}
\centering
\begin{tabular}{llrrr}
\hline\hline
 Asteroid & Year & Technique & PI \\
\hline
  (7889) 1994 LX        & 2022   & LC & Warner \\
  (68063) 2000 YJ66     & 2014   & LC & Warner \\
  (164121) 2003 YT1     & 2004   & Radar & Nolan\\
  (348400) 2005 JF21    & 2015   & Radar & Stephens \\
  (357439) 2004 BL86    & 2015   & LC    & Pollock \\ 
  (450894) 2008 BT18    & 2008   & Radar & Benner \\
  (523775) 2014 YB35    & 2015   & Radar & Naidu \\

\hline
\end{tabular}

\label{tab:VtypeNEA}
\end{table}

\subsection{Photometric set}

While spectroscopic classification is highly consistent and more nuanced than other methods of taxonomy, the availability of asteroid spectra severely limits its reach.

\begin{figure}[t]
\centering

\input{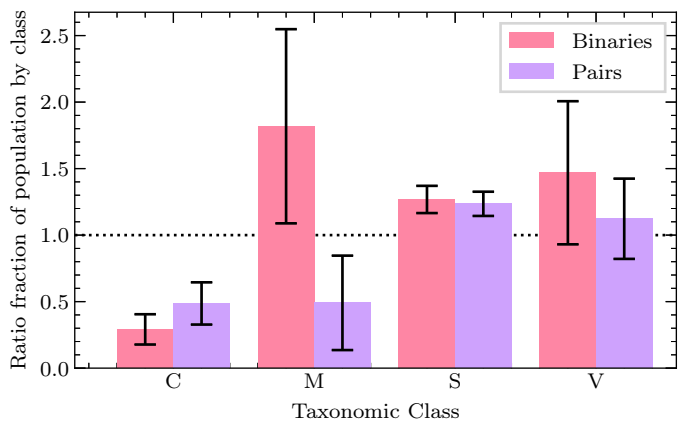}
\caption{Taxonomic distribution of binary asteroids relative
  to the general population, as measured from photometry.
  Q types are merged with S types in most photometric studies, and
  thus not separately available.
  }
\label{fig:photomresults}
\end{figure}

Although photometric classification schemes can be inconsistent,
the number of objects with a photometric classification is significantly
larger \KM{than} that of those with a spectroscopic classification,
with recent photometric studies such as
\citet{2018A&A...617A..12P},
\cite{2021A&A...652A..59S}, and
\cite{2022A&A...658A.109S}
containing over 200,000 objects,
while the spectroscopic set of \cite{2022A&A...665A..26M} contains only
around \numb{4000} asteroids.
As such, a much larger of number of binary and pair systems have been classified
in this manner. Because of this, we study these objects using the
technique described for the spectroscopic set, see
\Autoref{fig:photomresults}.

We find a significant under-representation of C complex asteroids
amongst binary systems, a significant over-representation of S types, \KM{and a slight over-representation of M complex asteroids, which are dominated by L type asteroids\footnote{\KM{In this case, the M complex contains no M type asteroids.}}}.
We find similar results amongst the pairs, with an additional
under-representation of M complex asteroids.


\section{Discussion\label{sec:discussion}}

The results presented in \Autoref{fig:spectraresults} and \Autoref{fig:photomresults}
point to strong differences in the abundance of taxonomic types amongst binary systems compared to the reference population.

We considered the C complex (encompassing C, B, P, and D classes) asteroids to be primitive,
due to a shared link between their opaque-rich surface compositions with that of inter-planetary
dust particles (IDPs) \citep{vernazza2015}. \KM{Although P types share spectral similarities to E, M and X type asteroids, and as such are often associated together as an X complex \citep{bus2002, busdemeotax}, this does not represent a true similarity in composition, and is therefore irrelevant for the purposes of this study.}
These objects also share characteristics with other remnants of the early Solar System, such as comets \citep{vbbook2016}.
We consider an additional super-complex of mafic-silicate rich asteroids, belonging to S, Q, V, and A,
as there is evidence for substantial heating in their formation history,
sourced from the geologic history of meteorite analogues \citep{1970mccordspectra, vernazza2014}.
K, L, M and X classes were not included in either category as these objects are of ambiguous origin
\citep[CO/CV and iron meteorites have been proposed as analogs,
e.g.,][]{sunshine2008,ockert-bell2010,2021Icar..35414034E},
and are not substantially opaque or mafic-silicate rich.
X type asteroids encompass objects with no known albedo that could belong to the
E, M or P classes, all of which have similar spectral properties but very different compositions
\citep{bus2002, busdemeotax, vernazza2015, 2022A&A...665A..26M}.
M type asteroids have been found to be metal rich, implying
that they are from thermally differentiated parent bodies, however, some members of this class
have been found to have signatures of silicates or hydration,
making the origin of these asteroids ambiguous \citep{Rivkin2000, 2022A&A...665A..26M}.

We compared the relative abundance of asteroids in these categories for the binaries and pairs with respect to their corresponding background populations.

For the spectroscopic set, there was an under abundance of \numb{$\Prmsig \sigma$} for primitive type asteroids, and an over abundance of \numb{$\Evlsig{} \sigma $} for S/Q/V/A type asteroids, suggesting that there is a strong preference for binary formation amongst S, Q, V and A type asteroids, with respect to primitive type asteroids.

\begin{figure}[t]
\centering
\input{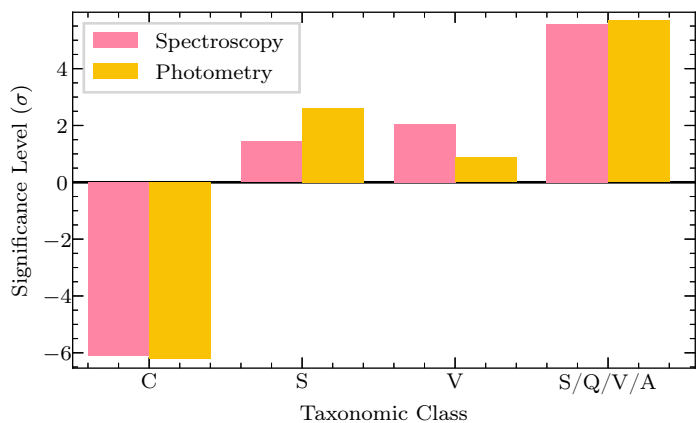}
\caption{Number of standard deviation away from the background for selected reduced taxonomic classes, with a negative value representing classes with lower representation in the specified population than the background, and a positive value representing a higher representation in the specified population than the background. Classes are order from least to most thermally evolved moving from left to right, with the most substantial under representation in the C complex, and the most substantial over representation in the V class and S/Q/V/A complex.}
\label{fig:significance_levels}
\end{figure}

Furthermore, in analysis of the reduced taxonomic classes
(\Autoref{fig:spectraresults}), it seems that the more
thermally evolved the parent body of an asteroid,
the more likely it is to form a binary system.
This trend can be seen in \Autoref{fig:significance_levels},
where the clear increase in standard deviations away
from agreement with the background can be seen as
the taxonomic groups become increasingly thermally evolved.
There is a strong under abundance in the C complex,
containing \KM{primitive IDPs},
and a strong over abundance of V type binary systems,
which are associated with the fully differentiated
parent body (4) Vesta. This increase is seen across
both the photometric and spectroscopic sets.

These findings are in alignment with recent theoretical models based on Hayabusa2
and OSIRIS-REx data by \cite{2022NatCo..13.4589Z}, which suggest that rapidly
rotating asteroids with primitive compositions are more likely to undergo
internal deformation than to form binary systems.
Our compilation of taxonomic classes for binary systems and creation of a
reference population supports these theoretical predictions.


\section{Conclusions\label{sec:conclusions}}

By collecting the taxonomic class for \numb{\nbBins} binary asteroids with
SsODNet \citep{berthier2022}, we find significant observational evidence for a non-random distribution
of binary systems amongst taxonomic classes.
There is an over-representation of multiple asteroids amongst mafic-silicate rich types (Q, V, A, E)
and under representation amongst primitive opaque-rich types (C, B, P, D).
The high representation of binary systems amongst V type NEAs also suggests that objects forming from
further differentiated parent bodies may be more likely to form binaries,
but limited sample size leaves this inconclusive.
While taxonomies from spectra are valued more highly in this study due to the
availability of a low error, highly consistent data set from \cite{2022A&A...665A..26M},
taxonomies from photometric measurements provide similar results.
The taxonomies of \numb{\nbPairs} asteroid pairs were also considered, based on the primary body of the pairs,
and we found that this taxonomic distribution is consistent with that found
in the set of binary systems.
This is in alignment with recent predictions by \citet{2022NatCo..13.4589Z}.
The consistency between the distributions of binaries and pairs implies that low-albedo
taxonomic types have difficulty forming binary systems, rather than difficulty maintaining them.

\begin{acknowledgement}
  We thank CNES for supporting K.\ Minker in her internship
  during her Masters of astrophysics
  (MAUCA: \url{http://mauca.oca.eu}).
  This research used the VO TOPCAT software
  \citep{2005ASPC..347...29T}.
  The properties of Solar System Object are from
  the service SsODNet.ssoCard of the
  Service des éphémérides de l'IMCCE through its
  Solar System Portal (\url{https://ssp.imcce.fr}).
  We thank all the developers and maintainers of
  resources for the community.
\end{acknowledgement}

\bibliographystyle{aa} 
\bibliography{biblio,ssodnet} 

\begin{appendix}

\pgfplotsset{compat=1.9}

\section{Compiled parameters\label{app:data_sources}}

The majority of data considered in this study was sourced from SsODNet \citep{berthier2022}, which contains a compilation of many independent studies. The exact parameters and objects considered in this study are available electronically at CDS. The first twelve lines of each set are included in this \Autoref{app:data_sources}. Column headings are as follows, `Number' (asteroid number), `Name' (asteroid name), `D' (primary diameter, if available), `$\sigma_D$' (uncertainty in primary diameter, if available), `D ref' (reference corresponding to primary diameter, if available), `$p_V$' (albedo, if available), `$\sigma_{p_V}$' (uncertainty in albedo, if available), `Tax. method' (method for taxonomy, either spectroscopy (spec) or photomotery (phot)), `Tax. scheme' (taxonomic scheme),  `Class' (taxonomic class), `Waverange' (waverange over which the taxonomy is determined), `Tax. Ref' (reference for the determination of the objects taxonomy). All albedo values are sourced from \cite{berthier2022},
who computed albedo from most-recent absolute magnitudes and best-estimates of diameter,
using \Autoref{eq:diameter}.

\begin{sidewaystable}[t]
  \caption{Parameters of included objects in the spectroscopic binaries set, first twelve lines included.}
\centering

\begin{tabular}{llrrrrrrrrrr}
	\toprule
	Number & Name & D & $\sigma_D$ & D Ref. & $p_V$ & $\sigma_{p_V}$ & Tax. method & Tax. scheme & Class & Waverange & Tax. Ref. \\
	\midrule
\hline \\1016 & Anitra & 9.98 & 0.07 & {\cite{2011ApJ...741...68M}} &  0.3 & 0.18  & Spec & Mahlke & S & VISNIR & {\cite{2022A&A...665A..26M}} \\
1798 & Watts & 6.88 & 0.09 & {\cite{2012ApJ...759L...8M}} &  0.24 & 0.19  & Spec & Mahlke & S & VISNIR & {\cite{2022A&A...665A..26M}} \\
2491 & Tvashtri & 3.13 & 0.09 & {\cite{2022PSJ.....3...30M}} &  0.7 & 0.19  & Spec & Mahlke & A & VISNIR & {\cite{2022A&A...665A..26M}} \\
2881 & Meiden & 5.84 & 0.07 & {\cite{2014ApJ...791..121M}} &  0.27 & 0.19  & Spec & Mahlke & S & VISNIR & {\cite{2022A&A...665A..26M}} \\
3122 & Florence & 4.4 & 0.03 & {\cite{2022PSJ.....3...30M}} &  0.26 & 0.18  & Spec & Mahlke & S & VISNIR & {\cite{2022A&A...665A..26M}} \\
3792 & Preston & 5.13 & 0.2 & {\cite{2016AJ....152...63N}} &  0.29 & 0.2  & Spec & Mahlke & S & VISNIR & {\cite{2022A&A...665A..26M}} \\
4435 & Holt & 5.56 & 0.27 & {\cite{2022PSJ.....3...56H}} &  0.26 & 0.21  & Spec & Mahlke & S & VISNIR & {\cite{2022A&A...665A..26M}} \\
4666 & Dietz & 6.73 & 0.4 & {\cite{2016AJ....152...63N}} &  0.25 & 0.22  & Spec & Mahlke & A & VISNIR & {\cite{2022A&A...665A..26M}} \\
7002 & Bronshten & 3.12 & 0.31 & {\cite{2017A&A...603A..55A}} &  0.2 & 0.27  & Spec & Mahlke & S & VISNIR & {\cite{2022A&A...665A..26M}} \\
15745 & Yuliya & 0.77 & 0.15 & {\cite{2010AJ....140..770T}} &  0.32 & 0.44  & Spec & Mahlke & S & VISNIR & {\cite{2022A&A...665A..26M}} \\
190166 & 2005 UP156 & 1.04 & 0.02 & {\cite{2015ApJ...814..117N}} &  0.2 & 0.19  & Spec & Mahlke & S & VISNIR & {\cite{2022A&A...665A..26M}} \\
226514 & 2003 UX34 & - & - & - & - & - & Spec & Mahlke & S & VISNIR & {\cite{2022A&A...665A..26M}} \\
\hline \\	\bottomrule
\end{tabular}

\label{tab:spectroscopicbins}
\end{sidewaystable}

\begin{sidewaystable}[t]
  \caption{Parameters of included objects in the photometric binaries set, first twelve lines included. }
\centering

\begin{tabular}{llrrrrrrrrrr}
	\toprule
	Number & Name & D & $\sigma_D$ & D Ref. & $p_V$ & $\sigma_{p_V}$ & Tax. method & Tax. scheme & Class & Waverange & Tax. Ref. \\
	\midrule
\hline \\2602 & Moore & - & - & - & - & - & Phot & Bus-DeMeo & S & VIS & {\cite{2013Icar..226..723D}} \\
2883 & Barabashov & 4.96 & 0.07 & {\cite{2014ApJ...791..121M}} &  0.32 & 0.19  & Phot & Bus-DeMeo & S & VIS & {\cite{2013Icar..226..723D}} \\
5500 & Twilley & 4.26 & 0.11 & {\cite{2022PSJ.....3...30M}} &  0.3 & 0.19  & Phot & Bus-DeMeo & S & VIS & {\cite{2013Icar..226..723D}} \\
6245 & Ikufumi & 7.88 & 0.21 & {\cite{2016AJ....152...63N}} &  0.13 & 0.19  & Phot & Bus-DeMeo & X & VIS & {\cite{2021A&A...652A..59S}} \\
7393 & Luginbuhl & 5.43 & 0.1 & {\cite{2015ApJ...814..117N}} &  0.27 & 0.19  & Phot & Bus-DeMeo & S & NIR & {\cite{2018A&A...617A..12P}} \\
9474 & Cassadrury & 3.57 & 0.16 & {\cite{2011ApJ...741...68M}} &  0.24 & 0.21  & Phot & Bus-DeMeo & S & VIS & {\cite{2013Icar..226..723D}} \\
18303 & 1980 PU & - & - & - & - & - & Phot & Bus-DeMeo & S & VIS & {\cite{2022A&A...658A.109S}} \\
20882 & Paulsanchez & - & - & - & - & - & Phot & Bus-DeMeo & S & VIS & {\cite{2013Icar..226..723D}} \\
25021 & Nischaykumar & 2.15 & 0.52 & {\cite{2011ApJ...741...68M}} &  0.16 & 0.52  & Phot & Bus-DeMeo & L & VIS & {\cite{2022A&A...658A.109S}} \\
26420 & 1999 XL103 & - & - & - & - & - & Phot & Bus-DeMeo & V & VIS & {\cite{2021A&A...652A..59S}} \\
72036 & 2000 XM44 & 3.1 & 0.58 & {\cite{2016AJ....152...63N}} &  0.28 & 0.42  & Phot & Bus-DeMeo & S & VIS & {\cite{2022A&A...658A.109S}} \\
250162 & 2002 TY57 & - & - & - & - & - & Phot & Bus-DeMeo & S & VIS & {\cite{2016Icar..268..340C}} \\
\hline \\	\bottomrule
\end{tabular}

\label{tab:photometricbins}
\end{sidewaystable}

\begin{sidewaystable}[t]
  \caption{Parameters of included objects in the photometric pairs set, first twelve lines included.}
\centering

\begin{tabular}{llrrrrrrrrrr}
	\toprule
	Number & Name & D & $\sigma_D$ & D Ref. & $p_V$ & $\sigma_{p_V}$ & Tax. method & Tax. scheme & Class & Waverange & Tax. Ref. \\
	\midrule
\hline \\34380 & Pratikvangal & 2.2 & 0.55 & {\cite{2011ApJ...741...68M}} &  0.26 & 0.53  & Phot & Bus-DeMeo & S & VIS & {\cite{2013Icar..226..723D}} \\
38184 & 1999 KF & 2.0 & 0.37 & {\cite{2011ApJ...741...68M}} &  0.31 & 0.41  & Phot & Bus-DeMeo & S & VIS & {\cite{2013Icar..226..723D}} \\
43008 & 1999 UD31 & - & - & - & - & - & Phot & Bus-DeMeo & S & VIS & {\cite{2022A&A...658A.109S}} \\
44620 & 1999 RS43 & - & - & - & - & - & Phot & Bus-DeMeo & S & VIS & {\cite{2022A&A...658A.109S}} \\
48652 & 1995 VB & 2.25 & 0.23 & {\cite{2011ApJ...741...68M}} &  0.23 & 0.27  & Phot & Bus-DeMeo & S & VIS & {\cite{2013Icar..226..723D}} \\
49791 & 1999 XF31 & - & - & - & - & - & Phot & Bus-DeMeo & S & VIS & {\cite{2022A&A...658A.109S}} \\
51609 & 2001 HZ32 & 1.97 & 0.67 & {\cite{2011ApJ...741...68M}} &  0.3 & 0.71  & Phot & Bus-DeMeo & V & VIS & {\cite{2022A&A...658A.109S}} \\
51866 & 2001 PH3 & 3.86 & 0.12 & {\cite{2015ApJ...814..117N}} &  0.25 & 0.19  & Phot & Bus-DeMeo & S & VIS & {\cite{2022A&A...658A.109S}} \\
52478 & 1995 TO & - & - & - & - & - & Phot & Bus & S & VIS & {\cite{2010A&A...510A..43C}} \\
55764 & 1992 DG12 & - & - & - & - & - & Phot & Bus-DeMeo & A & VIS & {\cite{2022A&A...658A.109S}} \\
55913 & 1998 FL12 & - & - & - & - & - & Phot & Bus-DeMeo & X & VIS & {\cite{2022A&A...658A.109S}} \\
56700 & 2000 LL28 & - & - & - & - & - & Phot & Bus-DeMeo & S & VIS & {\cite{2013Icar..226..723D}} \\
\hline \\	\bottomrule
\end{tabular}

\label{tab:photometricpairs}
\end{sidewaystable}

\begin{sidewaystable}[t]
  \caption{Parameters of included objects in the spectroscopic pairs set, first twelve lines included.}
\centering

\begin{tabular}{llrrrrrrrrrr}
	\toprule
	Number & Name & D & $\sigma_D$ & D Ref. & $p_V$ & $\sigma_{p_V}$ & Tax. method & Tax. scheme & Class & Waverange & Tax. Ref. \\
	\midrule
\hline \\42946 & 1999 TU95 & 4.75 & 0.1 & {\cite{2011ApJ...741...68M}} &  0.2 & 0.19  & Spec & Mahlke & S & VISNIR & {\cite{2022A&A...665A..26M}} \\
52852 & 1998 RB75 & 2.53 & 0.38 & {\cite{2011ApJ...741...68M}} &  0.27 & 0.35  & Spec & Mahlke & V & VISNIR & {\cite{2022A&A...665A..26M}} \\
53537 & Zhangyun & - & - & - & - & - & Spec & Mahlke & S & VISNIR & {\cite{2022A&A...665A..26M}} \\
54041 & 2000 GQ113 & 2.68 & 0.73 & {\cite{2011ApJ...741...68M}} &  0.28 & 0.57  & Spec & Mahlke & V & VISNIR & {\cite{2022A&A...665A..26M}} \\
54827 & Kurpfalz & 2.09 & 0.54 & {\cite{2011ApJ...741...68M}} &  0.24 & 0.55  & Spec & Mahlke & Q & VISNIR & {\cite{2022A&A...665A..26M}} \\
63440 & Rozek & - & - & - & - & - & Spec & Mahlke & C & VISNIR & {\cite{2022A&A...665A..26M}} \\
74096 & 1998 QD15 & - & - & - & - & - & Spec & Mahlke & S & VISNIR & {\cite{2022A&A...665A..26M}} \\
88604 & 2001 QH293 & 5.8 & 0.13 & {\cite{2011ApJ...741...68M}} &  0.2 & 0.19  & Spec & Mahlke & S & VISNIR & {\cite{2022A&A...665A..26M}} \\
92652 & 2000 QX36 & - & - & - & - & - & Spec & Mahlke & S & VISNIR & {\cite{2022A&A...665A..26M}} \\
101703 & 1999 CA150 & - & - & - & - & - & Spec & Mahlke & S & VISNIR & {\cite{2022A&A...665A..26M}} \\
\hline \\	\bottomrule
\end{tabular}

\label{tab:spectrscopicpairs}
\end{sidewaystable}

\begin{sidewaystable}[t]
  \caption{Parameters of included objects in the photometric binaries reference set, first twelve lines included.}
\centering

\begin{tabular}{llrrrrrrrrrr}
	\toprule
	Number & Name & D & $\sigma_D$ & D Ref. & $p_V$ & $\sigma_{p_V}$ & Tax. method & Tax. scheme & Class & Waverange & Tax. Ref. \\
	\midrule
\hline \\26920 & 1996 TQ12 & 1.52 & 0.2 & {\cite{2011ApJ...741...68M}} &  0.42 & 0.33  & Phot & Bus-DeMeo & Q & VIS & {\cite{2021A&A...652A..59S}} \\
30577 & 2001 OU103 & 1.7 & 0.2 & {\cite{2011ApJ...741...68M}} &  0.27 & 0.29  & Phot & Bus-DeMeo & S & VIS & {\cite{2022A&A...658A.109S}} \\
30958 & 1994 TV3 & 1.93 & 0.1 & {\cite{2011ApJ...741...68M}} &  0.63 & 0.21  & Phot & Bus-DeMeo & C & VIS & {\cite{2013Icar..226..723D}} \\
53428 & 1999 TD2 & - & - & - & - & - & Phot & Bus-DeMeo & X & VIS & {\cite{2022A&A...658A.109S}} \\
53440 & 1999 XQ33 & - & - & - & - & - & Phot & Bus & A & VIS & {\cite{2010A&A...510A..43C}} \\
55913 & 1998 FL12 & - & - & - & - & - & Phot & Bus-DeMeo & X & VIS & {\cite{2022A&A...658A.109S}} \\
63260 & 2001 CN & - & - & - & - & - & Phot & Bus-DeMeo & X & VIS & {\cite{2021A&A...652A..59S}} \\
74590 & 1999 OG2 & - & - & - & - & - & Phot & Bus-DeMeo & S & VIS & {\cite{2022A&A...658A.109S}} \\
82074 & 2000 YE119 & 2.69 & 0.19 & {\cite{2011ApJ...741...68M}} &  0.22 & 0.23  & Phot & Bus-DeMeo & S & VIS & {\cite{2013Icar..226..723D}} \\
85563 & 1998 BF7 & - & - & - & - & - & Phot & Bus-DeMeo & X & VIS & {\cite{2013Icar..226..723D}} \\
90216 & 2003 AS85 & - & - & - & - & - & Phot & Bus-DeMeo & L & VIS & {\cite{2022A&A...658A.109S}} \\
92036 & 1999 VZ180 & - & - & - & - & - & Phot & Bus-DeMeo & X & VIS & {\cite{2021A&A...652A..59S}} \\
\hline \\	\bottomrule
\end{tabular}

\label{tab:photometricbinsbk}
\end{sidewaystable}

\begin{sidewaystable}[t]
  \caption{Parameters of included objects in the spectroscopic binaries reference set, first twelve lines included.}
\centering

\begin{tabular}{llrrrrrrrrrr}
	\toprule
	Number & Name & D & $\sigma_D$ & D Ref. & $p_V$ & $\sigma_{p_V}$ & Tax. method & Tax. scheme & Class & Waverange & Tax. Ref. \\
	\midrule
\hline \\4581 & Asclepius & - & - & - & - & - & Spec & Mahlke & X & VISNIR & {\cite{2022A&A...665A..26M}} \\
33342 & 1998 WT24 & 0.4 & 0.06 & {\cite{2004PhDT.......371D}} &  0.67 & 0.35  & Spec & Mahlke & K & VISNIR & {\cite{2022A&A...665A..26M}} \\
88254 & 2001 FM129 & 0.8 & 0.01 & {\cite{2014ApJ...784..110M}} &  0.22 & 0.19  & Spec & Mahlke & Q & VISNIR & {\cite{2022A&A...665A..26M}} \\
138971 & 2001 CB21 & 0.34 & 0.01 & {\cite{2011PASJ...63.1117U}} &  0.58 & 0.2  & Spec & Mahlke & S & VISNIR & {\cite{2022A&A...665A..26M}} \\
152563 & 1992 BF & 0.37 & 0.1 & {\cite{2012ApJ...760L..12M}} &  0.16 & 0.56  & Spec & Mahlke & K & VISNIR & {\cite{2022A&A...665A..26M}} \\
164202 & 2004 EW & 0.16 & 0.03 & {\cite{2011AJ....141..109M}} &  0.31 & 0.37  & Spec & Mahlke & E & VISNIR & {\cite{2022A&A...665A..26M}} \\
208023 & 1999 AQ10 & 0.15 & 0.04 & {\cite{2016AJ....152..172T}} &  0.5 & 0.56  & Spec & Mahlke & S & VISNIR & {\cite{2022A&A...665A..26M}} \\
281375 & 2008 JV19 & - & - & - & - & - & Spec & Mahlke & X & VISNIR & {\cite{2022A&A...665A..26M}} \\
310442 & 2000 CH59 & - & - & - & - & - & Spec & Mahlke & Q & VISNIR & {\cite{2022A&A...665A..26M}} \\
401885 & 2001 RV17 & - & - & - & - & - & Spec & Mahlke & S & VISNIR & {\cite{2022A&A...665A..26M}} \\
474163 & 1999 SO5 & - & - & - & - & - & Spec & Mahlke & S & VISNIR & {\cite{2022A&A...665A..26M}} \\
518810 & 2010 CF19 & 0.08 & 0.01 & {\cite{2014ApJ...784..110M}} &  0.47 & 0.23  & Spec & Mahlke & S & VISNIR & {\cite{2022A&A...665A..26M}} \\
\hline \\	\bottomrule
\end{tabular}

\label{tab:spectroscopicbinsbk}
\end{sidewaystable}

\begin{sidewaystable}[t]
  \caption{Parameters of included objects in the photometric pairs reference set, first twelve lines included. }
\centering

\begin{tabular}{llrrrrrrrrrr}
	\toprule
	Number & Name & D & $\sigma_D$ & D Ref. & $p_V$ & $\sigma_{p_V}$ & Tax. method & Tax. scheme & Class & Waverange & Tax. Ref. \\
	\midrule
\hline \\163678 & 2002 XT65 & - & - & - & - & - & Phot & Bus-DeMeo & X & VIS & {\cite{2021A&A...652A..59S}} \\
177647 & 2004 RF110 & - & - & - & - & - & Phot & Bus-DeMeo & S & VIS & {\cite{2022A&A...658A.109S}} \\
182310 & 2001 OO43 & - & - & - & - & - & Phot & Bus-DeMeo & X & VIS & {\cite{2021A&A...652A..59S}} \\
191919 & 2005 QS157 & - & - & - & - & - & Phot & Bus-DeMeo & S & VIS & {\cite{2013Icar..226..723D}} \\
194922 & 2002 AK125 & - & - & - & - & - & Phot & Bus-DeMeo & B & VIS & {\cite{2022A&A...658A.109S}} \\
218019 & 2001 YX14 & - & - & - & - & - & Phot & Bus-DeMeo & S & VIS & {\cite{2022A&A...658A.109S}} \\
258019 & 2001 FY138 & - & - & - & - & - & Phot & Bus-DeMeo & S & VIS & {\cite{2022A&A...658A.109S}} \\
471448 & 2011 UZ145 & - & - & - & - & - & Phot & Bus-DeMeo & C & VIS & {\cite{2022A&A...658A.109S}} \\
475868 & 2007 CZ12 & - & - & - & - & - & Phot & Bus-DeMeo & X & VIS & {\cite{2022A&A...658A.109S}} \\
482128 & 2010 RY1 & - & - & - & - & - & Phot & Bus-DeMeo & V & VIS & {\cite{2022A&A...658A.109S}} \\
503415 & 2016 DC2 & - & - & - & - & - & Phot & Bus-DeMeo & X & VIS & {\cite{2021A&A...652A..59S}} \\
514044 & 2014 MM50 & - & - & - & - & - & Phot & Bus-DeMeo & X & VIS & {\cite{2021A&A...652A..59S}} \\
\hline \\	\bottomrule
\end{tabular}

\label{tab:photometricpairsbk}
\end{sidewaystable}

\begin{sidewaystable}[t]
  \caption{Parameters of included objects in the spectroscopic pairs reference set, first twelve lines included. }
\centering

\begin{tabular}{llrrrrrrrrrr}
	\toprule
	Number & Name & D & $\sigma_D$ & D Ref. & $p_V$ & $\sigma_{p_V}$ & Tax. method & Tax. scheme & Class & Waverange & Tax. Ref. \\
	\midrule
\hline \\1920 & Sarmiento & 2.85 & 0.16 & {\cite{2011ApJ...741...68M}} &  0.51 & 0.22  & Spec & Mahlke & E & VISNIR & {\cite{2022A&A...665A..26M}} \\
4736 & Johnwood & 2.72 & 0.07 & {\cite{2022PSJ.....3...30M}} &  0.78 & 0.19  & Spec & Mahlke & E & VISNIR & {\cite{2022A&A...665A..26M}} \\
6461 & Adam & 2.6 & 0.47 & {\cite{2016AJ....152...63N}} &  0.43 & 0.41  & Spec & Mahlke & E & VISNIR & {\cite{2022A&A...665A..26M}} \\
7829 & Jaroff & 2.73 & 0.16 & {\cite{2011ApJ...741...68M}} &  0.44 & 0.22  & Spec & Mahlke & E & VISNIR & {\cite{2022A&A...665A..26M}} \\
15692 & 1984 RA & 1.72 & 0.32 & {\cite{2011ApJ...741...68M}} &  0.58 & 0.41  & Spec & Mahlke & E & VISNIR & {\cite{2022A&A...665A..26M}} \\
16585 & 1992 QR & 2.0 & 0.29 & {\cite{2011ApJ...741...68M}} &  0.57 & 0.35  & Spec & Mahlke & E & VISNIR & {\cite{2022A&A...665A..26M}} \\
20043 & Ellenmacarthur & - & - & - & - & - & Spec & Mahlke & S & VISNIR & {\cite{2022A&A...665A..26M}} \\
25884 & Asai & 1.91 & 0.46 & {\cite{2011ApJ...741...68M}} &  0.6 & 0.52  & Spec & Mahlke & E & VISNIR & {\cite{2022A&A...665A..26M}} \\
53424 & 1999 SC3 & - & - & - & - & - & Spec & Mahlke & S & VISNIR & {\cite{2022A&A...665A..26M}} \\
138970 & 2001 CV19 & 1.09 & 0.11 & {\cite{2017A&A...603A..55A}} &  0.28 & 0.27  & Spec & Mahlke & S & VISNIR & {\cite{2022A&A...665A..26M}} \\
230269 & 2001 XZ6 & - & - & - & - & - & Spec & Mahlke & S & VISNIR & {\cite{2022A&A...665A..26M}} \\
275611 & 1999 XX262 & 1.48 & 0.01 & {\cite{2020PSJ.....1....5M}} &  0.04 & 0.18  & Spec & Mahlke & Ch & VISNIR & {\cite{2022A&A...665A..26M}} \\
\hline \\	\bottomrule
\end{tabular}

\label{tab:spectrscopicpairsbk}
\end{sidewaystable}

\end{appendix}

\end{document}